\documentclass[11pt]{article}
\usepackage{amssymb}
\usepackage{amsmath}
\usepackage{amscd}
\usepackage{latexsym}

\catcode`@=11
\renewcommand{\section}{\@startsection{section}{1}{0pt}{\medskipamount}
{\medskipamount}{\large\bf}}
\numberwithin{equation}{section}
\catcode`@=12

\def\b{\beta}
\def\g{\gamma}

\def\de{\delta}
\def\ve{\varepsilon}

\def\th{\theta}

\def\m{\mu}
\def\n{\nu}

\def\s{\sigma}
\def\p{\phi}
\def\vp{\varphi}

\def\1{\bar 1}
\def\2{\bar 2}
\def\3{\bar 3}
\def\Ups{\Upsilon}

\newcommand{\yb}{\bar{y}}

\newcommand{\C}{\mathbb C}
\newcommand{\R}{\mathbb R}
\newcommand{\T}{\mathbb T}

\newcommand{\Acal}{{\cal A}}
\newcommand{\Ncal}{{\cal N}}

\newcommand{\Fcal}{{\cal F}}
\newcommand{\Ecal}{{\cal E}}

\def\im{\mbox{i}}
\def\N2{$N{=}2$}
\def\pa{\mbox{$\partial$}}
\def\diff{\mbox{d}}
\def\tr{{\rm tr}}
\def\sfrac#1#2{{\textstyle\frac{#1}{#2}}}
\def\>{\rangle}
\def\<{\langle}
\def\+{\dagger}
\def\={\ =\ }

\begin{document}
\begin{titlepage}
\setcounter{page}{0}
\begin{flushright}
April 2008\\
ITP-UH-12/08
\end{flushright}

\vskip 1.5cm

\begin{center}

{\Large\bf Bounces/Dyons in the Plane Wave Matrix Model
\\[6pt]
and SU($N$) Yang-Mills Theory}

\vspace{12mm}
{\large Alexander D. Popov$^*$}
\\[3mm]
\noindent {\em Institut f\"ur Theoretische Physik,
Leibniz Universit\"at Hannover \\
Appelstra\ss{}e 2, 30167 Hannover, Germany }
\\
{Email: {\tt popov@itp.uni-hannover.de}}
\\[2mm]
and
\\[2mm]
\noindent {\em Bogoliubov Laboratory of Theoretical Physics, JINR\\
141980 Dubna, Moscow Region, Russia}
\\
{Email: {\tt popov@theor.jinr.ru}}
\vspace{12mm}

\begin{abstract}
\noindent We consider SU($N$) Yang-Mills theory on the space $\R\times S^3$
with Minkowski signature $(-+++)$. The condition of SO(4)-invariance imposed
on gauge fields yields a bosonic matrix model which is a consistent truncation
of the plane wave matrix model. For matrices parametrized by a scalar $\phi$,
the Yang-Mills equations are reduced to the equation of a particle moving in
the double-well potential. The classical solution is a bounce, i.e. a particle
which begins at the saddle point $\phi=0$ of the potential, bounces off the
potential wall and returns to $\phi=0$. The gauge field tensor components
parametrized by $\phi$ are smooth and for finite time both electric
and magnetic fields are nonvanishing. The energy density of this
non-Abelian dyon configuration does not depend on coordinates of
$\R\times S^3$ and the total energy is proportional to the inverse radius
of $S^3$. We also describe similar bounce dyon solutions in SU($N$) Yang-Mills
theory on the space $\R\times S^2$ with signature $(-++)$. Their energy is
proportional to the square of the inverse radius of $S^2$. From the
viewpoint of Yang-Mills theory on $\R^{1,1}\times S^2$ these solutions
describe non-Abelian (dyonic) flux tubes extended along the $x^3$-axis.

\end{abstract}
\end{center}

\vfill

\textwidth 6.5truein
\hrule width 5.cm

{\small
\noindent ${}^*$
Supported in part by the Deutsche Forschungsgemeinschaft (DFG).}

\end{titlepage}

\section{Introduction}

\noindent
The idea of AdS/CFT correspondence~\cite{1,2,3} is one of the most
important concepts in studying nonperturbative aspects of string,
gravity and gauge theories. In particular, it was recently shown that
large $N$ weakly coupled SU($N$) gauge theory on $S^1\times S^3$
demonstrates a confinement-deconfinement transition at temperatures $T$
proportional to the inverse radius of $S^3$ and the corresponding black
hole-string transition was discussed (see~\cite{4,5,6} and references therein).
Note that for finite radius $R$ of $S^3$ and nonzero temperature $T=1/L$,
where $L$ is the circumference of $S^1$, there are finite-action
non-BPS solutions of Yang-Mills equations on $S^1\times S^3$ describing
$n$ instanton-antiinstanton pairs~\cite{7}. Furthermore, they exist only
if $L\ge L_n=2\pi Rn$, i.e. for $T\le T_n=1/(2\pi Rn)$ or $TR\le 1/(2\pi n)$.
Maybe it is not coincidence that the number $n$ of such instanton-antiinstanton
pairs increases for $T\to 0$; in other words, $n\to\infty$ if $T\to 0$.

Recently, new proposals for gauge/gravity correspondence were formulated~\cite{8}.
In particular, the Lin-Maldacena method works for the plane wave matrix
model~\cite{9}, for maximally supersymmetric Yang-Mills (SYM) theory on
$\R\times S^2$~\cite{8,10} and for $\Ncal =4$ SYM theory on $\R\times S^3/Z_k$
on the gauge side (see~\cite{11}-\cite{13} and references therein). Using this
method, one can describe smooth gravity solutions corresponding to vacua of
the above SYM models and study the gravity description of instantons interpolating
between gauge vacua~\cite{8,11,12}.

Instantons in the plane wave matrix model and SU($N$) Yang-Mills theory
on $\R\times S^2$ were described e.g. in~\cite{14,7}. It would be of interest
to construct some other explicit solutions in these models for further checking
nonperturbative aspects of gauge/gravity correspondence. In this paper, we
describe finite-energy dyon configurations
in Yang-Mills theories on the spaces $\R\times S^3$ and $\R\times S^2$ with
Minkowski signature. In fact, imposing SO(4)-invariance we reduce Yang-Mills
theory on $\R\times S^3$ to a bosonic matrix model which is the $\Ncal =0$
subsector of the $\Ncal =4$ plane wave matrix model in conformity with the
previous results~\cite{15,16,17}. We briefly discuss perspectives of integrability
of $\Ncal =4$ SYM and plane wave matrix theories. Then we reduce Yang-Mills
theory on $\R\times S^2$ to a non-Abelian (matrix) analog of the $\phi^4$
kink model.
In the simplest case the solution of both matrix models is a bounce, i.e.
a particle in the double-well potential which begins at $\phi =0$ for
$t=-\infty$, bounces off the potential wall at $t=0$ and returns to $\phi =0$
for $t=+\infty$ (cf.~\cite{18}). In this case, matrices in the above
models are constant matrices multiplied by $\phi$, and we obtain finite-energy
dyon solutions of Yang-Mills equations on $\R\times S^3$ and $\R\times S^2$.
Furthermore, solutions on $\R\times S^2$ can be uplifted to configurations
with finite energy (dyons) on the space $\R\times S^2\times S^1$ and to
configurations with finite energy per unit length along $x^3$-axis
(vortex tube) on the space $\R\times S^2\times \R$.

\section{Dyon configurations in Yang-Mills theory on $\R\times S^3$}

\noindent
{\bf Manifold $\R\times S^3$.}  Let us consider the space $\R\times S^3$ with
Minkowski signature $(-+++)$. On the standard three-sphere $S^3$ of the constant
radius $R$ we consider one-forms\footnote{See e.g.~\cite{12,7} for the
explicit expression.} $\{e^a\}$ satisfying the Maurer-Cartan equations
\begin{equation}\label{2.1}
\diff e^a - \frac{1}{\sqrt{2}R}\, \ve^a_{bc}\, e^b\wedge e^c =0\ ,
\end{equation}
where $a,b,...=1,2,3$. Introducing $e^0:=\diff x^0 =\diff t$, we can write the
metric on $\R\times S^3$ in the form
\begin{equation}\label{2.2}
\diff s^2 = -(e^0)^2 + \de_{ab}\, e^a\, e^b \ .
\end{equation}
Note that our choice of $e^a$'s differs from the standard one by the factor
$\sqrt{2}$. The standard choice can be restored e.g. by rescaling
$R$ in (\ref{2.1}) and other formulae.

\smallskip

\noindent
{\bf SO(4)-invariance.} On $\R\times S^3$ we consider a gauge potential $\Acal$
and the gauge field $\Fcal =\diff\Acal + \Acal\wedge\Acal$ taking values in the
Lie algebra $su(N)$. Since $S^3=\,$SO(4)/SO(3) is a homogeneous SO(4)-space,
we can impose on $\Acal$ and $\Fcal$ a condition of SO(4)-invariance defined up to
gauge transformations (cf.~\cite{19}) which in the `temporal gauge' $\Acal_t =0$
yields
\begin{equation}\label{2.3}
\Acal = \sfrac12\, X_ae^a\ ,\quad
\Fcal = \sfrac{1}{2}\, \dot X_a\, \diff t \wedge e^a +
\sfrac{1}{2}\, \bigl (\sfrac{1}{\sqrt{2}R}\,\ve^c_{ab}\, X_c +
\sfrac{1}{4}\, [X_a, X_b]\bigr )\, e^a\wedge e^b\ ,
\end{equation}
where the overdot denotes differentiation with respect to time.

Substitution of (\ref{2.3}) into the standard Yang-Mills equations on
$\R\times S^3$ leads to the second order equations on $X_a$:
\begin{equation}\label{2.4}
\ddot X_a+\frac{2}{R^2}X_a+\frac{3}{2\sqrt{2}R}\, \ve_{abc}\, [X_b, X_c]+
\frac{1}{4}\,\bigl [X_b, [X_a,X_b]\bigr]=0\ .
\end{equation}
These equations describe a bosonic matrix model which is a consistent
truncation from $\Ncal =4$ to $\Ncal =0$ of the plane wave matrix model~\cite{9}.
In particular, (\ref{2.4}) coincide with the proper equations in~\cite{16}
after the redefinition $X_a\mapsto -2\im X_a$ and $R\mapsto R/\sqrt{2}$.
For more details on the reduction of $\Ncal =4$ SYM theory to the plane
wave matrix model and further studies see e.g.~\cite{15,16,17,12}.

\smallskip

\noindent
{\bf Bounces in the plane wave matrix model.} Let us consider the ansatz
(cf.~\cite{7})
\begin{equation}\label{2.5}
X_a = \Bigl (\phi - \frac{\sqrt{2}}{R}\Bigr ) \,T_a\ ,
\end{equation}
where $\phi =\phi(t )$ is a real-valued function of $t$ and $T_a$'s are generators
of $N$-dimensional representation of SU(2). Substituting (\ref{2.5}) into (\ref{2.4}),
we obtain the equation
\begin{equation}\label{2.6}
\ddot\phi - \frac{1}{R^2}\phi + \frac{1}{2}\phi^3=0
\end{equation}
which can also be obtained as the Euler-Lagrange equation from the reduced
Yang-Mills action after integration over $S^3$. This action describes a
particle with the kinetic energy density
\begin{equation}\label{2.7}
T(\phi )=\dot\phi^2
\end{equation}
moving in the double-well potential
\begin{equation}\label{2.8}
U(\phi )=\frac{1}{4}\Bigl(\frac{2}{R^2}-\phi^2 \Bigr)^2\ .
\end{equation}

The solution of eq.~(\ref{2.6}) is known as a bounce (see e.g.~\cite{18}
and references therein)
\begin{equation}\label{2.9}
\phi =\frac{2}{R\cosh (\frac{t}{R})}\quad\Rightarrow\quad
\dot\phi =-\frac{2\sinh (\frac{t}{R})}{R^2\cosh^2 (\frac{t}{R})}
\end{equation}
since
\begin{equation}\label{2.10}
\phi (\pm\infty )=0\ , \quad \phi (0)=\frac{2}{R}\ ,\quad
\dot\phi (\pm\infty )=0\quad\mbox{and}\quad \dot\phi (0)=0\ ,
\end{equation}
i.e. the particle starts at the saddle point $\phi=0$ of the potential
(\ref{2.8}), bounces off the potential wall on the right at $\phi =\frac{2}{R}$,
and returns to $\phi=0$ at $t=+\infty$. Of course, in (\ref{2.9}) one can shift
$t\mapsto t-t_0$ due to translational invariance.

For the solution (\ref{2.9}) we have
\begin{equation}\label{2.11}
T(\phi )=\frac{4\sinh^2 (\frac{t}{R})}{R^4\cosh^4 (\frac{t}{R})}\ ,\quad
U(\phi )=\frac{(\cosh^2 (\frac{t}{R})-2)^2}{R^4\cosh^4 (\frac{t}{R})}
\end{equation}
and the energy density is
\begin{equation}\label{2.12}
\Ecal (\phi )=T(\phi ) + U(\phi )=\frac{1}{R^4}\ .
\end{equation}

\smallskip

\noindent
{\bf Dyons in Yang-Mills theory.} Substituting (\ref{2.9}) into (\ref{2.3}),
we obtain a dyon configuration
\begin{subequations}\label{2.13}
\begin{eqnarray}
\Acal &=& \frac{1}{\sqrt{2}R}\Bigl (\frac{\sqrt{2}}{\cosh(\frac{t}{R})}-1\Bigr )\,
T_a\,e^a\ ,\\
\Fcal &=& -\frac{\sinh (\frac{t}{R})}{R^2\cosh^2 (\frac{t}{R})}\,T_c\,\diff t\wedge
e^c + \frac{1}{4R^2}\Bigl (\frac{2}{\cosh^2(\frac{t}{R})}-1\Bigr)\,\ve^c_{ab}
\,T_c\, e^a\wedge e^b\ .
\end{eqnarray}
\end{subequations}
In fact, we have
\begin{equation}\label{2.14}
E_a=-\frac{\sinh (\frac{t}{R})}{R^2\cosh^2 (\frac{t}{R})}\,T_a
\quad\mbox{and}\quad
B_a=\frac{1}{2R^2}\Bigl (\frac{2}{\cosh^2(\frac{t}{R})}-1\Bigr)\,T_a
\end{equation}
for the non-Abelian electric and magnetic fields $E_a=E_a^c\, T_c=\Fcal_{0a}$
and $B_a=B_a^c\,T_c=\sfrac12\,\ve_{abc}\,\Fcal_{bc}$. Note that
$E_a\to 0$ and $B_a\to -\sfrac{1}{2R^2}\,T_a$ for $t\to\pm\infty$, i.e.
the asymptotic fields are nonvanishing and on the 3-spheres $S^3$ at
$t = \pm\infty$ we have
\begin{equation}\label{2.15}
 \Acal  = - \frac{1}{\sqrt{2}R}\, e^a T_a = \frac{1}{2}\, g^{-1} \diff g\ ,
\end{equation}
where $g$ is a smooth map $g: S^3 \to$ SU($N$) of winding number one.

For the energy density of the bounce dyon configuration we obtain
\begin{equation}\label{2.16}
\Ecal = -\tr (E_aE_a+B_aB_a)=\frac{1}{16R^4}\sum\limits_k p_k(p_k^2-1)
\le\frac{1}{16R^4}N(N^2-1)\ ,
\end{equation}
where it is assumed that $T_a$'s are generators of reducible $N$-dimensional
representation $(p_1,...,p_K )$ of SU(2) such that $\sum_kp_k=N$. For
irreducible representation we simply have $p_1=N$. For the energy we have
\begin{equation}\label{2.17}
E= \int_{S^3}\diff^3x\,\sqrt{g}\, \Ecal=\frac{\pi^2}{8R}\sum\limits_k p_k(p_k^2-1)
\end{equation}
with Vol$(S^3)=2\pi^2R^3$. Note that the dependence on $p_k$'s in (\ref{2.16}),
(\ref{2.17}) will disappear for another usual choice of $T_a$'s such
that $\tr (T_aT_b)=-\sfrac12\de_{ab}$.

\smallskip

\noindent
{\bf Towards integrability of the plane wave matrix model.} For the matrix
model (\ref{2.4}) the instanton subsector is described by the first order
BPS equations (see e.g.~\cite{14,7}) with $t\mapsto\tau =\im\, t$.\footnote{It
is of interest to generalize noncommutative instantons on $\R^4$ (see
e.g.~\cite{20} and references therein) to the space $\R\times S^3$. Note
that one can consider a quantum group type deformation of $\R\times S^3$
since this space is a Lie group.}
These first order equations are transformed to the standard Nahm equations
by the redefinitions
\begin{equation}\label{2.18}
X_a\mapsto Y_a=\frac{1}{2} \exp\Bigl(-\frac{\sqrt{2}}{R}\tau\Bigr)
X_a\quad\mbox{and}\quad
\tau\mapsto r=\frac{R}{\sqrt{2}}\exp\Bigl(\frac{\sqrt{2}}{R}\tau\Bigr)\quad
\Rightarrow\quad \tau =\frac{R}{\sqrt{2}}\log\Bigl(\frac{\sqrt{2}}{R}r\Bigr)
\end{equation}
discussed in~\cite{21} along with their integrability, Lax pair, charges and
the finite-dimensional moduli space. Note that Nahm's equations can be algebraically
reduced to the (periodic) Toda chain equations (see e.g.~\cite{22}), solutions
of which are known explicitly. In fact, the integrability of the BPS
subsector of the model (\ref{2.4}) follows from the integrability of the
self-dual Yang-Mills equations on $\R\times S^3$ after imposing SO(4)-invariance.
Furthermore, the redefinitions in (\ref{2.18}) correspond to the well-known conformal
transformations from the space $\R\times S^3$ to the space $\R^4\setminus \{0\}$,
\begin{equation}\label{2.19}
\diff\tau^2 +\frac{R^2}{2}\diff\Omega^2_3=\frac{R^2}{2r^2}(\diff r^2+
r^2\diff\Omega^2_3)\ ,
\end{equation}
so that translations in $\tau$ (and $t$) correspond to dilatations in
$\R^4\setminus \{0\}$ (see e.g.~\cite{15,16} and references therein).

The second order equations (\ref{2.4}), corresponding to the SO(4)-invariant
subsector of the full Yang-Mills theory on $\R\times S^3$, are not integrable.
However, it is well known that the $\Ncal =3$ SYM equations (equivalent to $\Ncal =4$
ones) in Minkowski signature can be represented in the twistor approach as the
compatibility conditions of some linear equations on an auxiliary function
$\psi$~\cite{23} (see~\cite{24} for recent reviews and references). In fact,
this function $\psi$ depending on an extra `spectral' parameter encodes all
the information about the $\Ncal =4$ SYM multiplet and the above `zero curvature'
representation hints on integrability of $\Ncal =4$ SYM theory and of the plane
wave matrix model which is its SO(4)-reduction. For deriving the system of linear
equations for the matrix model one should write them down on the $\Ncal =3$
superambitwistor space for $\R\times S^3$ and impose the condition of
SO(4)-equivariance on all fields. It is expected that, similar to the case of
vortex equations on Riemann surfaces of genus $g>1$~\cite{25}, these linear
differential equations on $\psi$ will keep derivatives along $S^3$ in spite of the
fact that the plane wave matrix model is formulated in $0+1$ dimensions. Note that
we are speaking about {\it nonperturbative} integrability which might stem from `Lax
representation' of nonlinear equations of motion and not about the perturbative one
discussed e.g. in~\cite{17}.

\section{Dyons in Yang-Mills theory on $\R\times S^2$}

Here we want to construct bouncing dyon configurations in Yang-Mills theory
on $\R\times S^2$ similar to the case of gauge theory on $\R\times S^3$.
Namely, imposing the condition of SO(3)-invariance, we reduce the Yang-Mills
equations on $\R\times S^2$ to bosonic matrix equations in $0+1$ dimensions.
Note that this matrix model can be obtained via some limit and truncation
from the plane wave matrix model as discussed e.g. in~\cite{8,11,12}. For
discussion of gravity dual to maximally supersymmetric Yang-Mills theory
on $\R\times S^2$ and $\R\times S^3$ see~\cite{8},\cite{11}-\cite{13} and
references therein. However, we will not
discuss these correspondences here. Instead, we embedd the model
into Yang-Mills theory in $3+1$ dimensions and describe its dyon solutions.

\smallskip

\noindent
{\bf Manifold $\R\times S^2\times\T$.} We consider the space $\R\times S^2\times\T$
with Minkowski signature $(-+++)$, where $\T$ is $S^1$ or $\R$. We choose local
real coordinates $x^\mu$ with indices $\mu ,\nu ,...$ running through 0,1,2,3
so that $x^1$, $x^2$ are local coordinates on $S^2$ and $x^3$ is a coordinate
on $\T$. We also denote by $x^i$ the coordinates $x^0, x^3$ and introduce
on $\R\times\T$ the Minkowski metric $\eta =(\eta_{ij})= \mbox{diag}(-1,+1)$
with $i,j,...=0,3$. On $S^2\cong\C P^1$ we introduce the local complex coordinate
$y=\sfrac12(x^1+\im x^2)$ related with angle coordinates $0\le\th <\pi$,
$0\le\vp\le 2\pi$ by
\begin{equation}\label{3.1}
y=R\tan\Bigl(\frac{\th}{2}\Bigr)\exp(-\im\vp )\quad\mbox{and}\quad
\yb=R\tan\Bigl(\frac{\th}{2}\Bigr)\exp(\im\vp )\ ,
\end{equation}
where bar denotes complex conjugation. In these coordinates
the metric on $\R\times S^2\times \T$ has the form
\begin{eqnarray}\label{3.2}
\diff s^2&=&g_{\m\n}\diff x^\m\,\diff x^\n =
\diff s^2_{\Sigma^{1,1}}+\diff s^2_{\C P^1}
= \eta_{ij} \diff x^i\diff x^j +2\,g_{y\yb}\,\diff y\,\diff \yb
=\\
&=&-(\diff x^0)^2 + (\diff x^3)^2+R^2(\diff\th^2+\sin^2\th\,\diff\vp^2)=
-(\diff x^0)^2 + (\diff x^3)^2+
\frac{4R^4}{(R^2+y\yb)^2}\,\diff y\,\diff \yb\ ,\nonumber
\end{eqnarray}
where $\Sigma^{1,1}=\R\times\T$.

\smallskip

\noindent
{\bf SO(3)-invariance.} The question of invariance of the gauge fields on
manifolds $X\times S^2$ under the action of the isometry group
SO(3)$\,\cong\,$SU(2) of $S^2$ was discussed e.g. in~\cite{26,27}. Here we restrict
ourselves to a particular SU(2)-invariant ansatz for fields on
$\Sigma^{1,1}\times S^2$ described in~\cite{7}. It has the form
\begin{equation}\label{3.3}
\Acal = a\Ups_m + \sfrac12\, \Phi_m\bar\b - \sfrac12\, \Phi_m^\+\b\ ,
\end{equation}
where
\begin{equation}\label{3.4}
a= \frac{1}{2(R^2+y\yb)}(\yb\,\diff y - y\,\diff\yb )\ ,\quad
\b =\frac{{\sqrt 2}\,R^2}{R^2+y\yb}\, \diff y\ ,
\end{equation}
\begin{equation}\label{3.5}
\Ups_m=\mbox{diag} (m,...,m-2\ell ,..., -m)\quad\mbox{for}\quad\ell =0,...,m
\end{equation}
and
\begin{equation}\label{3.6}
\Phi_m=\begin{pmatrix}0&\phi_1&...&0\\\vdots&0&\ddots&\vdots\\
\vdots&&\ddots&\phi_m\\0&...&...&0\end{pmatrix}\ .
\end{equation}
Here $\phi_\ell =\bar\phi_\ell$ for $\ell = 1,...,m$ are real scalar fields,
$a$ in (\ref{3.3}) and (\ref{3.4}) is the gauge potential on the Dirac
one-monopole line bundle over $\C P^1$ and $\b$ is the (1,0) type
form on $\C P^1$.

For the gauge field tensor components we have
\begin{subequations}\label{3.7}
\begin{eqnarray}
\Fcal_{ij}&=& 0\ ,\quad
\Fcal_{y\yb}=- \frac{1}{4}g_{y\yb} \bigl (\frac{2}{R^2}\Ups_m -
[\Phi_m,\Phi^\+_m]\bigr )\ ,\\
\Fcal_{i\yb}&=&\sfrac12\, {\rho}\,\pa_i\Phi_m \quad\mbox{and}\quad
\Fcal_{iy}=-\sfrac12\,{\rho}\,\pa_i\Phi_m^\+ \ ,
\end{eqnarray}
\end{subequations}
where
\begin{equation}\label{3.8}
\rho =(g_{y\yb})^{1/2}=\frac{4\sqrt{2}\, R^2}{4R^2+(x^1)^2+(x^2)^2}\ .
\end{equation}

\smallskip

\noindent
{\bf Matrix $\Phi^4$ type model.}
Substituting (\ref{3.3})-(\ref{3.7}) into the Yang-Mills action on
$M:=\R\times S^2\times\T$ and integrating over $\C P^1$, we obtain
\begin{equation}\label{3.9}
S=-\frac{1}{4\pi}{\int_M}\tr(\Fcal{\wedge}*\Fcal )=
R^2{\int_{\Sigma^{1,1}}}\diff^2x\,\tr\Bigl\{\pa_i\Phi_m\pa^i\Phi_m^\+
+\frac{1}{8}\bigl (\frac{2}{R^2}\Ups_m -[\Phi_m,\Phi_m^\+]\bigr )^2\Bigr\}\ ,
\end{equation}
where $*$ is the Hodge operator. From (\ref{3.9}) we obtain the matrix field
equations
\begin{equation}\label{3.10}
\pa_i\pa^i\Phi_m + \frac{1}{R^2}\Phi_m -
\frac{1}{4}\bigl [[\Phi_m,\Phi_m^\+], \Phi_m\bigr]=0\ ,
\end{equation}
which is equivalent to the linked equations
\begin{equation}\label{3.11}
\pa_i\pa^i\p_\ell + \frac{1}{R^2}\p_\ell + \frac{1}{4}\bigl(\p^2_{\ell -1}-2\p^2_\ell +
\p^2_{\ell +1}\bigr) \p_\ell=0
\end{equation}
with $\ell =1,...,m$ and $\p_0:=0=:\p_{m+1}$.

\smallskip

\noindent
{\bf $\Phi^4$ bounces.} We impose the condition $\pa_3\p_\ell =0$ reducing
(\ref{3.7})-(\ref{3.11}) to the Yang-Mills model on the space
$\R\times S^2$ with Minkowski signature $(-++)$. From (\ref{3.11}) we obtain
the equations
\begin{equation}\label{3.14}
\ddot\p_\ell - \frac{1}{R^2}\p_\ell - \frac{1}{4}\bigl(\p^2_{\ell -1}-2\p^2_\ell +
\p^2_{\ell +1}\bigr) \p_\ell=0\ .
\end{equation}
As solution of these equations we have
\begin{equation}\label{3.15}
\Phi_m=\frac{2}{R\cosh (\frac{t}{R})}\Phi_m^0\ ,
\end{equation}
where $\Phi_m^0$ are given by
\begin{equation}\label{3.13}
\Phi^0_m\ :\quad \p^0_\ell
= \pm\sqrt{\ell (m-\ell +1)}\quad\mbox{for}\quad \ell =1,...,m\ ,
\end{equation}
so that $\frac{\sqrt{2}}{R}\,\Phi^0_m$ are the vacua of the model.

As more general SU($N$) solutions we can take
\begin{subequations}\label{3.16}
\begin{eqnarray}
m=2r&:&\p_1=\frac{2}{R\cosh (\frac{t-a_1}{R})}\ , \quad \p_2=0\ , \quad
\p_3=\frac{2}{R\cosh (\frac{t-a_3}{R})}\ ,
\nonumber\\
&&\quad \ldots\quad
\p_{m-1}=\frac{2}{R\cosh (\frac{t-a_{m-1}}{R})}\ , \quad
\p_m=0\\
m=2r+1&:&\p_1=\frac{2}{R\cosh (\frac{t-a_1}{R})}\ ,\quad
 \p_2=0\ , \quad
\p_3=\frac{2}{R\cosh (\frac{t-a_3}{R})}\ ,
\nonumber\\
&&\quad \ldots\quad \p_{m-1}=0 \ , \quad
\p_{m}=\frac{2}{R\cosh (\frac{t-a_m}{R})}\ ,
\end{eqnarray}
\end{subequations}
where each $\p_\ell\ne 0$ describes a bounce with different moduli $a_\ell$.

\smallskip

\noindent
{\bf Dyons.} To obtain dyon configurations in SU($N$) Yang-Mills theory
we should substitute the bounce solution (\ref{3.15}) or (\ref{3.16}) into
(\ref{3.3})-(\ref{3.7}). Note that from the non-Abelian bounces in $0+1$
dimensions we obtain dyon solutions of Yang-Mills equations in $3+1$ dimensions,
on $\R\times S^2\times\T$, similar to monopoles on the same space obtained
from $\Phi^4$ kinks~\cite{7}. Furthermore, in
$3+1$ dimensions we can compare them by applying to both a Lorenz rotation,
substituting $t\mapsto\g (t-vx^3)$ for bounces and $x^3\mapsto\g (x^3-vt)$
for kinks, where $\g = (1-v^2)^{-1/2}$ and $-1<v<1$.

We will write down here the explicit form of $\Acal$ and $\Fcal$ only for
the $su(2)$-bounce (\ref{3.15}) with $m=1$. Namely, substituting (\ref{3.15})
into (\ref{3.3}) and (\ref{3.7}), we obtain
\begin{equation}\label{3.17}
\Acal = a\s_3+\frac{1}{R\cosh (\frac{t}{R})}\bigl (\bar\b\s_+ - \b\s_-\bigr )
\quad\mbox{with}\quad\s_+=\begin{pmatrix}0&1\\0&0\end{pmatrix}=\s_-^\+\ ,
\end{equation}
\begin{equation}\label{3.18}
\Fcal = \frac{\sqrt{2}\sinh (\frac{t}{R})}{R^2\cosh^2 (\frac{t}{R})}\,\b^0\wedge
\Bigl (\b^1\frac{\s_2}{2\im} - \b^2\frac{\s_1}{2\im}\Bigr )
- \frac{1}{R^2}\Bigl (1-\frac{2}{\cosh^2(\frac{t}{R})}\Bigr)\,\frac{\s_3}{2\im}
\b^1\wedge \b^2\ ,
\end{equation}
where
\begin{equation}\label{3.19}
\b^0:=\diff t\ ,\quad \b^1:=\frac{1}{\sqrt{2}}(\b+\bar\b)\ ,\quad
\b^2:=-\frac{\im}{\sqrt{2}}(\b-\bar\b)\quad\mbox{and}\quad
\b^3:=\diff x^3
\end{equation}
form the nonholonomic basis of one-forms on $\R\times S^2\times\T$ such that
for (\ref{3.2}) we have
\begin{equation}\label{3.20}
\diff s^2= -(\b^0)^2+(\b^1)^2+(\b^2)^2+(\b^3)^2\ .
\end{equation}
Thus, from the viewpoint of $3+1$ dimensions we have
\begin{equation}\label{3.21}
E_1^2=\frac{\sqrt{2}\sinh (\frac{t}{R})}{R^2\cosh^2 (\frac{t}{R})}=-E_2^1\ ,
\quad E^a_3=0\quad\mbox{and}\quad B_1^a=0=B_2^a\ ,\quad
B_3^3=\frac{1}{R^2}\Bigl (\frac{2}{\cosh^2(\frac{t}{R})}-1\Bigr)
\end{equation}
with the energy density
\begin{equation}\label{3.22}
\Ecal = -\tr (E_aE_a+B_aB_a)=\frac{1}{2}(E_a^cE_a^c+B_a^cB_a^c)=\frac{1}{2R^4}\ .
\end{equation}
Therefore, in $2+1$ dimensions we have
\begin{equation}\label{3.23}
E_{S^2}=\int_{S^2}\diff^2x\,\sqrt{g}\ \Ecal =\frac{2\pi}{R^2}\ ,
\end{equation}
which can be considered as the energy density per unit length along the space $\T$.
For $\T=S^1$ we can integrate further obtaining
\begin{equation}\label{3.24}
E_{S^2\times S^1}=\frac{2\pi L}{R^2}\ ,
\end{equation}
where $L$ is the circumference of $S^1$. For $\T=\R$ we obtain the dyonic vortex
tube extended along the $x^3$-axis.

\section{Concluding remarks}

In the paper~\cite{7} we have shown how kinks in the $\phi^4$ type models in $1+0$
dimensions can be uplifted to instantons of Yang-Mills theories on $S^3\times\R$
and $S^2\times\R$. Similarly, sphalerons in the same $\phi^4$ type models uplifted
to chains of instanton-antiinstanton pairs on $S^3\times S^1$ and $S^2\times S^1$.
Note that instanton configurations in Yang-Mills theory on the space $S^2\times\T$
can be interpreted as static monopole configurations on $\R\times S^2\times\T$, where
$\T$ is $\R$ or $S^1$. In this paper, we have shown that bounces in $\phi^4$ type
models in $0+1$ dimensions can be uplifted to dyons of Yang-Mills theory on the
spaces $\R\times S^3$, $\R\times S^2$ and $\R\times S^2\times\T$ with Minkowski
signatures in each case. These dyons have finite energy densities on the spaces
$S^3\subset\R\times S^3$, $S^2\subset\R\times S^2$ and $S^2\times S^1
\subset\R\times S^2\times S^1$. In fact, dyons on $\R\times S^3$ are bounce solutions
of the plane wave matrix model and dyons in Yang-Mills theory on $\R\times S^2$
are bounce solutions of the related matrix model.

\smallskip

It would be of interest
\begin{itemize}
\item
to study gravity dual description of the above-mentioned monopole, dyon and
instanton configurations following~\cite{8,11,13}
\item
to construct supersymmetric generalizations of our exact solutions
\item
to consider quantum effects in the nonperturbative background defined by these
monopole, dyon and instanton configurations
\item
to construct noncommutative generalizations of the above-mentioned solutions
 to the Yang-Mills equations
\end{itemize}
As a more general perspective, it is of interest to study classical and quantum
 integrability of maximally supersymmetric Yang-Mills theories on the spaces
$\R\times S^2$ and $\R\times S^3$ and of the related matrix
models in $0+1$ dimensions.


\end{document}